# SUB-NANOSECOND ELECTRON EMISSION FROM ELECTRICALLY GATED FIELD EMITTING ARRAYS

**M. Paraliev[ξ], S. Tsujino, C. Gough, E. Kirk, S. Ivkovic**
*Paul Scherrer Institute*
*5232 Villigen PSI, Switzerland*

## *Abstract*

Field Emitting Arrays (FEAs) are a promising alternative to the conventional cathodes in different vacuum electronic devices such as traveling wave tubes, electron accelerators and etc. Electrical gating and modulation capabilities, together with the ability to produce stable and homogeneous electron beam in high electric field environment are the key requirements for their practical application. Due to relatively high gate capacitance, fast controlling of FEA emission is difficult. In order to achieve sub-nanosecond, electrically controlled, FEA based electron emission a special pulsed gate driver was developed. Bipolar high voltage (HV) pulses are used to rapidly inject and remove charge form FEA gate electrode controlling quickly electron extraction gate voltage. Short electron emission pulses (<600 ps FWHM) were observed in low and high gradient (up to 12 MV/m) environment. First attempts were made to combine FEA based electron emission with radio frequency acceleration structures (1.5 GHz) using pulsed preacceleration. The gate driver design together with low inductance FEA chip contact system is described. The results obtained in low and high gradient experimental setups are presented and discussed.

## I. INTRODUCTION

The advancement in fabrication of metallic Field Emitting Array (FEA) devices [1, 2] opened new possibilities for development of high current density cold cathodes, suitable for different electronic devices like microwave generators and amplifiers [3, 4], plasma sources [5], electron accelerators [6] etc. Direct electrical modulation of single gated FEAs was successfully demonstrated in traveling-wave tubes where the cathode current has been modulated in microwave frequency range [7, 8, 9]. For electron accelerators application, cathodes capable of producing short, high current electron bunches are required. Direct electrical gating of cathode current is an attractive feature of FEA based cold cathodes. In order to preserve electron bunch length and quality (avoid space charge degradation) the bunch has to be accelerated quickly up to relativistic energies. This sets the requirement that for such applications, FEA based cathodes should be able to operate stably and reliably in high electric gradients. Based on beam dynamics studies double gated FEA based cathode has bean proposed as an alternative high brilliance electron source [10]. As a first step towards realization of a practical FEA based accelerator cathode, we have explored fast current switching capabilities of single gated FEA devices as well as their compatibility with high gradient electric acceleration environment. A HV gate driver scheme was employed in order to produce sub-nanosecond long electron bunches. Encouraged by the achieved short switching times we explored the further RF acceleration of the FEA based bunches (up to 5 MeV) using Low Emittance Gun (LEG) test stand in Paul Scherrer Institut, Switzerland [11].

## II. FEA CONSTRUCTION AND OPERATION

The FEA structure is fabricated by molding and self-aligned gate process [12]. The emitter array (Fig. 1.) consists of pyramidal molybdenum tips (*b)* 1.2 µm high with square base 1.5 x 1.5 µm, supported by 0.4 mm thick nickel substrate (*a*). The gate electrode (*d*) consists of 0.5 µm thick molybdenum layer separated from the emitter substrate by 1.2 µm tick $SiO_2$ insulation layer (*c*). In the described experiments circular shape FEAs with diameter 2.26 mm were used. The number of tips and the tips' pitch in the tested arrays are respectively 120 x $10^3$ and 5 µm. Fig. 2 shows typical scanning electron microscope image of the FEA.

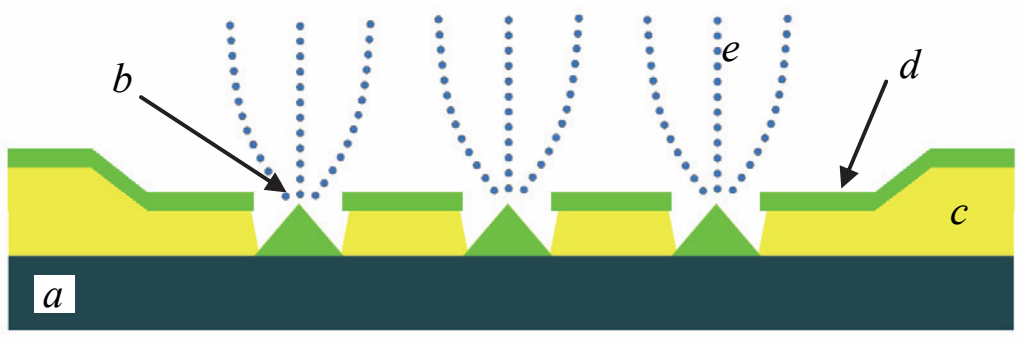

Fig. 1. Schematic cross section view of FEA emitter: *a* – nickel substrate, *b* – field emitting tip, *c* – $SiO_2$ insulation layer, *d* – gate electrode and *e* – emitted electrons.

[ξ] email: martin.paraliev@psi.ch

If negative voltage is applied between the emitting tips (substrate) and the gate electrode, due to emitter shape defined field enhancement factor a high electric field is established around the pyramids' apexes. The metal-vacuum potential barrier is narrowed and slightly lowered (Schottky lowering) allowing the electrons to tunnel through it.

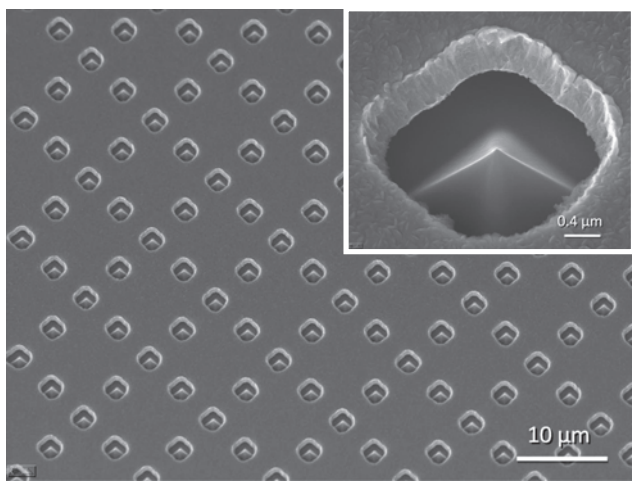

Fig. 2. Scanning electron microscope micrograph of FEA array (sample) and an individual pyramid emitter.

The amount of emitted current $I_{em}$ (eq. 1) is well described by the analytical model developed by R. Fowler and L. Nordheim [13] based on electron tunneling probability through thin potential barrier. Fig 3. compares a measured FEA emission characteristic with the theoretically predicted one. Because of the exponential current-voltage dependency, relatively small changes in the gate voltage $U_g$ will produce significant change in the emitted current. It could be easily seen that ~1/3 decrease of gate voltage will reduce the emitted current two orders of magnitude. Practically that means the field emission is shut off. Together with the fact that field emission is an instantaneous process it gives a good ground for fast on-off controlling of FEA devices.

$$I_{em} = aAF^2 e^{(-b\frac{\varphi^{3/2}}{F})}, \qquad (1)$$

where $a$ and $b$ are constants, $A$ is emitting area, $F = \beta Ug/d$ is extraction electric field, $d$ is gate to emitter distance and $\varphi$ is work function of the emitting material.

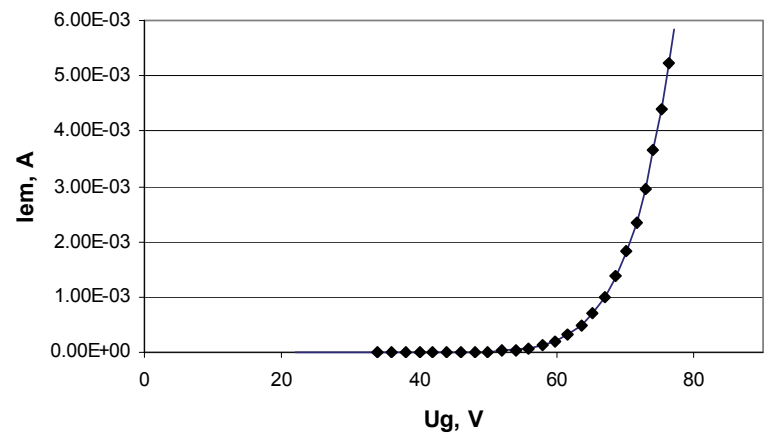

Fig. 3. FEA emission characteristic (points) compared with the analytical expression based on Fowler-Nordheim tunneling theory (line).

The lowest switching time limit is estimated taking in account the finite FEA dimensions and finite resistance of the gate layer (the substrate resistance is neglected because the substrate is much thicker than the gate). To switch the whole FEA on (or off) one should consider the propagation time of the radial electromagnetic wave from the periphery to the center of the emitting array trough the transmission line formed between the substrate and the gate electrode. Since the dielectric constant of $SiO_2$ is 3.8 the wave group velocity is 1.95 times lower than speed of light. For 2 mm diameter FEA the wave propagation time is in order of 7 ps. The front delay due to finite gate electrode resistance is numerically estimated to be in order of 20 ps (for the same FEA diameter). FEA switching in sub-nanosecond time scale should not be limited by these two values.

## III. CONTACT SYSTEM

The gate electrode capacitance of the produced FEAs was 1.3 nF. This represents a heavy capacitive load for the driver circuit. Moreover, the standard FEA chips connection method (bonding wires) introduces stray inductance in the driving circuit that limits the driving speed to tens of ns. In order to achieve sub-nanosecond long field emission two important topological modifications were done. First: the thickness of SiO2 isolation layer outside the active matrix was increased (roughly twice). This brought gate electrode capacitance down to 0.7 nF. Second: A polished metal lip ($c$) a of the FEA holder ($a$) is in direct contact with the front surface of the gate layer (Fig. 4).

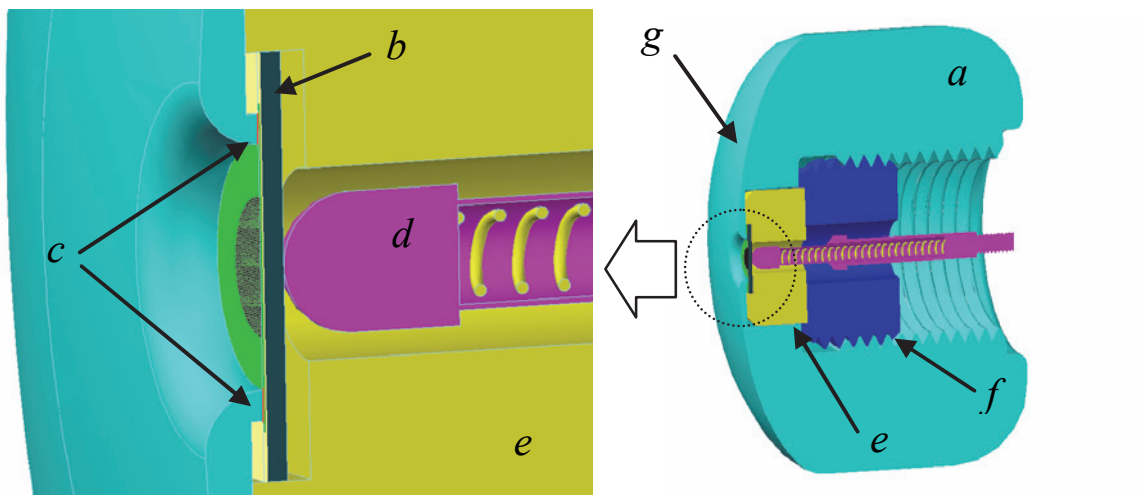

Fig. 4 Cross section of the cathode (left is a magnified view of the circled area): *a* – FEA holder, *b* – FEA chip, *c* – polished metal lip that contacts directly to FEA gate, *d* – spring loaded contact, *e* – ceramic nest, *f* – copper nut and *g* – DLC coated surface.

In order to protect the FEA chip ($b$) from excessive mechanical stress controlled force is applied to the FEA chip back side through spring loaded contact ($d$). A ceramic nest ($e$) defines FEA chip transverse position. The spring-loaded contact forms roughly matched continuation of the driving 50 Ohm transmission line (not shown on the figure). In this way high speed driving stimuli could be applied to the FEA structure. The FEA holder ($a$) is made out of stainless steel. Its surface is polished and Diamond Like Carbon (DLC) coated in order to withstand high surface electric field.

## IV. DRIVING SCHEME

In order to have short voltage pulses over the gate electrode capacitance it is needed to quickly inject and remove charge from it. Eq. 2 gives the amount of charge $Q$ needed for given voltage change $\Delta U_g$ over the gate electrode with capacitance $C_g$

$$Q = \Delta U_g C_g \quad (2)$$

Since for short pulses the gate impedance is much smaller than the transmission line characteristic impedance ($X_{G(1GHz)} = 0.2\ \Omega >> Z_L = 50\ \Omega$), the transmission line could be viewed as voltage controlled current source with coefficient 1/50. Two identical HV pulses, with opposite polarity are sent to the FEA gate through the transmission line. For low frequency (including DC) the gate is an open circuit. The fast controlling pulses are sent on top of a DC bias in order to reduce the needed voltage swing over the gate capacitance and to avoid excessive currents during charge injection and removal. Fig. 5 illustrates the driver (Udrv) and gate electrode (Ug) voltage waveforms.

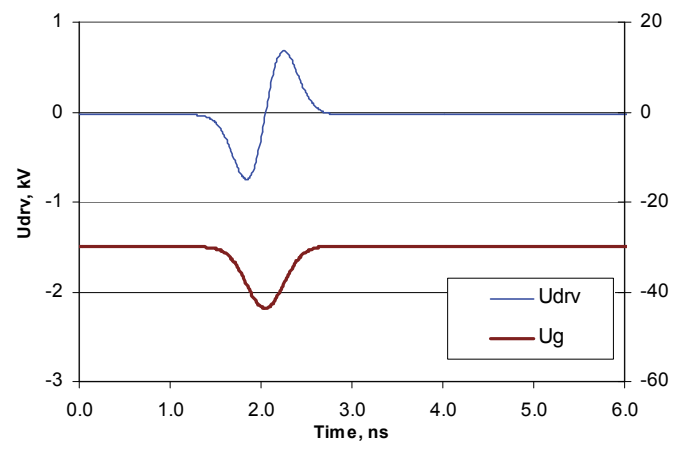

Fig. 5. Driver (Udrv) and gate (Ug) voltage waveforms.

Two different schemes were used to generate the needed bipolar double pulse waveform. In both cases a short pulse 5 kV generator was used (FPG 5-1PN from FID GmbH [14]).

The first scheme uses a shorted transmission line stub as it is shown in fig. 6 top). The short negative pulse splits in junction A where the stub T3 is connected to the main transmission line T1-T2. A negative pulse with 67% of the original pulse amplitude continues to the FEA. A second pulse with the same amplitude propagates in the stub. The end of the stub is AC shorted and when the pulse reaches the stub end reflects with opposite polarity (becoming positive). It propagates back in the stub until it reaches junction A and splits again. A pulse with 67% of the reflected pulse amplitude (or 44% of the original amplitude) and delayed with twice of the electrical length of the stub T3 is sent to the FEA. The emission pulse length is defined by the delay between the two pulses. A DC bias is applied through the stub AC termination.

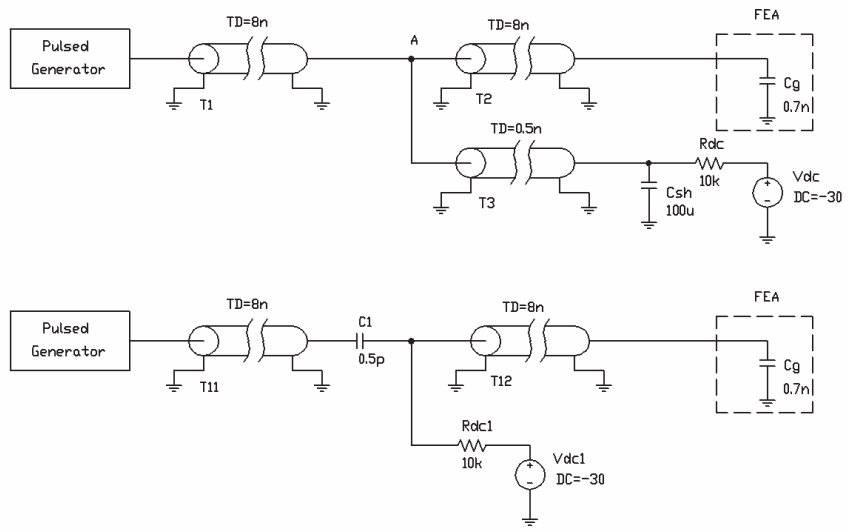

Fig. 6. Driving schemes: Top - "Shorted line" and bottom - "Differentiator".

The second (positive) pulse is 33% smaller amplitude than the first (negative) but it is large enough to stop the field emission. This scheme introduces less loss – up to 67% of the charge of the initial HV pulse is delivered to the FEA gate electrode. The drawback is that the shortest FEA gate pulse is limited down to the double the length of the original pulse. In the second scheme (Fig 6. bottom) a differentiator is used to convert the initial pulse in two consequent bipolar pulses. For a reasonable performance of a RC differentiator the impedance of the capacitor has to be more than 10 times higher than the resistor value (in this case the impedance of the transmission line). This requirement costs signal amplitude reduction with the same factor. The differentiating capacitor C1 is used as a DC decoupling to inject a DC bias voltage. The circuit is capable of "replicating" the original HV pulse shape (length) on the gate electrode and in this way gives the shortest possible pulse. The major drawback is the charge transfer ratio from the original HV pulse to the FEA gate electrode is in order of 10% - much less than in the first scheme. Fig. 7 shows the measured waveforms produced by the two driving schemes. The "long cable" waveforms take in account the dispersion in the 25 m long connecting cable for the 500 kV setup.

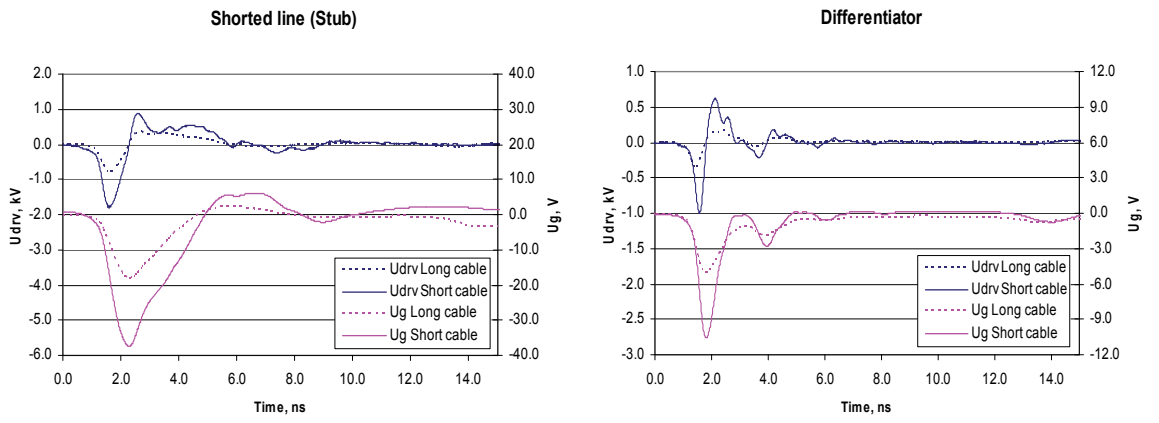

Fig. 7. Driving pulse (Udrv) and gate (Ug) waveforms: Left - "Shorted line" and right -"Differentiator".

## V. TEST SETUPS

Two setups were used to study the short pulse field emission. FEA holder and contact system are kept identical. In the first one the measurement vacuum diode consists of the described FEA holder system and a coaxial Faraday cup as an anode. The separation between them is 10 mm and the maximum anode voltage is limited to 4 kV due to the N type coaxial vacuum feedthrough used.

The second setup is the LEG test stand [11] that includes pulsed diode acceleration up to 500 keV followed by radiofrequency (RF) accelerating structures (max acceleration 5 MeV). The described above FEA holder is attached to 500 kV pulsed transformer and the anode is grounded. The FEA gate control signal is delivered through 25 m coaxial cable (the secondary of the pulsed transformer). The cable length and dispersion attenuate and degrade the driving waveforms. Anode-cathode gap and voltage are variable and it makes possible to explore different accelerating voltages and gradients.

## VI. TEST RESULTS

The first test setup was used to condition the FEAs and to observe directly the emitted charge using fast oscilloscope. Fig. 8 shows a family of registered electron pulses for different anode voltage. With lower accelerating voltages the short electron pulse degrades due to space charge forces. In this setup the maximum accelerating voltage and gradient are limited to 4 kV and respectively 0.4 MV/m. The shortest electron pulse from this family is 525 ps FWHM.

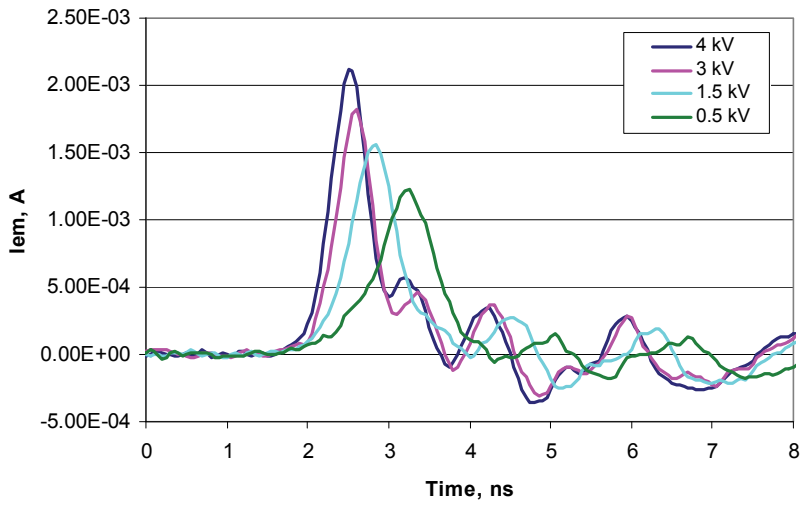

Fig. 8. Emitted electron pulses for different anode voltages with "differentiator" driving scheme in low gradient setup.

In the second setup the FEAs were tested in high gradient environment – up to 22 MV/m and pulsed pre-acceleration up 350 keV. The short electron pulses were further accelerated up to 5 MeV and the pulses' length was estimated using the modulation depth of an RF phase scan. Fig. 9. shows the collected charge vs RF phase.

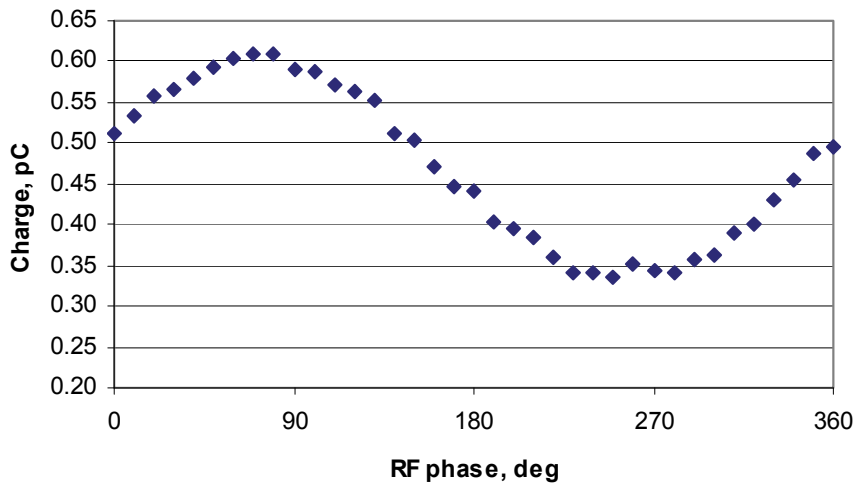

Fig. 9. Modulation of the collected charge in high gradient setup.

Modulation dept is 30.5% and corresponds to Gaussian pulse 420 ps FWHM wide (sigma 180 ps). If the jitter contribution is taken in account the width of the electron pulse is estimated to be below 400 ps. Fig. 10. shows YAG screen image of 0.6 pC, 3.5 MeV FEA generated electron beam. Taking in account the pulse duration emitted current is estimated to be 1.5 mA. The maximum achieved charge and current were respectively 2.4 pC and 6 mA.

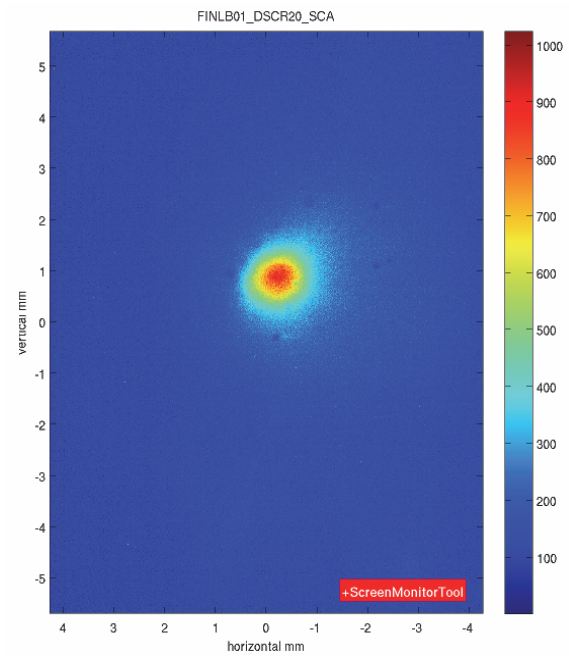

Fig. 10. Cold cathode electron beam image after RF acceleration structure (0.6 pC, 3.5 MeV)

# VII. SUMMARY

Fast FEA driver was developed using a pair of identical bipolar HV pulses capable of controlling the field emission in sub-nanosecond time scale. Low stray inductance FEA holder was used where the usual wire bonding contacts to the FEA chip were substituted by direct surface contact. The thickness of $SiO_2$ insulating layer in the area around the field emitting array was increased in order to reduce the parasitic FEA gate capacitance. Sub-nanosecond electrically gated field emission was successfully demonstrated. Electron pulses in order of 0.5 ns FWHM were studied in low and high accelerating gradient. Due to the short electron pulses it was possible to further accelerate those up to 5 MeV using double cell 1.5 GHz RF cavity. Different gas discharge treatments were explored as potential methods to further increase electron beam peak current and spatial homogeneity. Fast electrical control of single gated FEAs in high accelerating gradient environment was a successful step towards the practical implementation of double gated field emitting cathode capable of producing short high brilliance electron pulses.